\documentclass{article}

\usepackage[main,final]{neurips_2026}

\usepackage[utf8]{inputenc} 
\usepackage[T1]{fontenc}    
\usepackage{hyperref}       
\usepackage{url}            
\usepackage{booktabs}       
\usepackage{amsfonts}       
\usepackage{nicefrac}       
\usepackage{microtype}     
\usepackage{xcolor}         
\usepackage{mathtools}

\usepackage{amsmath,amsfonts, amssymb, amsthm}
\usepackage{thmtools}

\theoremstyle{definition}
\newtheorem{theorem}{Theorem}%[section]

\newtheorem{claim}[theorem]{Claim}
\newtheorem*{claim*}{Claim}
\newtheorem{lemma}[theorem]{Lemma}

\newcommand{\dist}{\text{dist}}
\newcommand{\diam}{\text{diam}}

\newcommand{\card}[1]{\vert #1 \vert}
\newcommand{\SP}{\text{sp}}

\newcommand{\reals}{\mathbb{R}}

\DeclareMathOperator*{\argmin}{\arg\!\min}

\newcommand{\E}{\mathbb{E}} % expected value
\newcommand{\Prob}{\mathbb{P}} % probability

\usepackage{graphicx}
\usepackage{subcaption}
\usepackage{caption}

\title{Markov Metrical Task Systems}

\author{%
  Christian Coester\thanks{Authors listed in alphabetical order.}\\
  Department of Computer Science\\
  University of Oxford\\
  United Kingdom\\
  \texttt{christian.coester@cs.ox.ac.uk}
  \And
  Alexa Tudose\footnotemark[1]\\
  Department of Computer Science\\
  University of Oxford\\
  United Kingdom\\
  \texttt{maria-alexa.tudose@cs.ox.ac.uk}
}

\begin{document}

\maketitle

\begin{abstract}
  We study Metrical Task Systems, a broad class of problems in online algorithms, under a model where input requests are generated by a stochastic process. We first consider the case of requests drawn IID from an unknown distribution, and then extend the model to requests generated by a finite-state Markov chain, where the algorithm observes both the current request and the current state of the chain before making its next move. This formulation captures practical settings in which requests are generated by an evolving environment. In both settings, we show how to find in polynomial time an algorithm whose expected cost per step is arbitrarily close to the cost of the best possible online algorithm. Our theoretical results are complemented by a brief empirical evaluation of our algorithms.
\end{abstract}

\section{Introduction} \label{sec:intro}
\emph{Metrical Task Systems (MTS)}, introduced by Borodin et al.\ \cite{MTSintroduction}, is a central problem in online decision-making.
In MTS, an algorithm controls a server moving in a metric space $\mathcal{M} = (M, d)$, starting from an initial state $s_0 \in M$. At each time step $t\geq 1$, a request (task) is revealed in the form of a cost function $c_t\colon M \rightarrow \reals_{\geq 0} \cup \{\infty\}$, specifying the cost of serving the request from each state. In response, the algorithm selects a state $s_t \in M$, pays cost $d(s_{t-1}, s_t)$ to move the server to $s_t$, and cost $c_t(s_t)$ to serve the request. The goal is to design an algorithm which minimizes the total cost incurred. Despite its simple definition, MTS captures a broad class of online problems as special cases, including caching~\cite{CompetitiveRatioDef}, $k$-server \cite{Manasse88}, layered graph traversal \cite{FiatLGT}, and convex body chasing \cite{Friedman1993}.

Traditionally, MTS has mostly been studied in the \emph{adversarial} model, where the request sequence is allowed to be arbitrary, and can even be chosen adversarially. The performance of an online algorithm is then evaluated based on its \emph{competitive ratio}, defined as the worst case ratio between algorithm's cost and the best solution in hindsight \cite{CompetitiveRatioDef}. This model is attractive because it makes no assumptions on the input, but this flexibility comes at the cost of pessimism. Even though strong impossibility results are known in the adversarial model, practical applications often do not resemble worst-case instances, as the requests are not arbitrary but instead follow an underlying pattern. 

In this work, we study MTS under \emph{stochastic requests}. We first consider the case where requests are drawn independently from an unknown distribution. We then consider the richer case where requests are generated by a finite-state Markov chain, and the online algorithm observes both the current request and the current Markov state before making its next move. This formulation captures many practical settings in which requests are generated by an evolving environment. For example, in caching, the MTS state represents the cache contents and Markov requests capture temporal locality in page accesses \cite{MarkovPaging, Lund99, NewMarkovPaging}. In mobility applications such as bike-share, the MTS state is a spatial allocation of resources, while Markov requests model temporal patterns such as commuting, nightlife, weekend demand or weather.

This stochastic formulation changes the nature of the problem. In the adversarial setting, the central question is how an online algorithm compares to the hindsight optimum on the worst possible instance. In our setting, the goal is to characterize and compute an optimal policy for a fixed environment. The resulting problem is related to average-cost Markov decision processes, as outlined in section \ref{sec:preliminaries}.

Formally, in the \emph{IID requests} model, we assume that each service cost function $c_t$ is drawn independently from an unknown distribution $\mathcal{D}$, and we assume access to a sampling oracle for $\mathcal{D}$. In the \emph{Markov requests} model, we assume that for a known finite set $\mathcal X$, for each $x \in \mathcal X$ there exists a distribution $\mathcal D_x$ over $\mathcal X\times [0,\infty]^M$ such that, conditioning on $x_{t-1} = x$, the pair $(x_t,c_t)$ is sampled from $\mathcal{D}_x$. Given an initial $x_0\in\mathcal X$, this induces a Markov chain $(x_t)_{t\ge 0}$, which we assume to be irreducible. The distributions $(\mathcal{D}_x)_{x \in \mathcal{X}}$ are not known in advance. Note that the IID case is a special case where $\card{\mathcal{X}} = 1$. We consider three ways of obtaining training data. With \emph{conditional oracles}, the learner chooses a state $x$ and
receives an independent pair $(x',c)\sim\mathcal D_x$. With a \emph{stationary oracle}, each call independently returns a
triple $(x,x',c)$ by first drawing $x\sim p$, where $p$ is the stationary distribution of the Markov chain, and then
$(x',c)\sim\mathcal D_x$. With an \emph{observed trajectory}, we are given a realization of 
Markov states $x_0, x_1, \dots, x_N$ and cost functions $c_1, \dots, c_N$.

\subsection{Our results} \label{sec:our-results}

In the cases of both IID requests and Markov requests, we show how to find in polynomial time a policy whose expected cost is arbitrarily close to the cost of the best possible \textit{online} algorithm. 

We denote by $g^*$ the average cost per step of the optimal online policy.
Also, we denote by $n$ the number of points in the underlying metric space and by $D$ its diameter.

\begin{theorem}[IID requests]\label{th:culminating-iid}
Let $\varepsilon \in (0,D)$ and
$\delta \in (0,1)$. By using
$O\left(
    \frac{D^2}{\varepsilon^2}
    \left(
        n \log \frac{D}{\varepsilon}
        +
        \log \frac{n}{\delta}
    \right)
\right)
$
independent samples drawn from $\mathcal D$, with probability at least $1-\delta$ we find
in polynomial time a policy with average cost at most $g^*+\varepsilon$ per step.
\end{theorem}

\begin{restatable}{theorem}{theoremMarkov}{(Markov requests with oracle access).}\label{th:culminating-Markov}
 Let $\varepsilon \in (0,D)$, $\delta \in (0,1)$.
With either a conditional oracle or a stationary oracle,
by using $O\left (\frac{D^2 \card{\mathcal{X}}^2}{{\varepsilon}^2} \left(n \log \frac{D}{\varepsilon}
        + \log \frac{n \vert \mathcal{X} \vert}{\delta}\right) \right)$ oracle calls, with probability at least $1-\delta$ we find in polynomial time a policy with average cost at most $g^* + \varepsilon$ per step.
\end{restatable}

\begin{restatable}{theorem}{theoremTrajectory}{(Markov requests with observed trajectory).}\label{th:culminating-Markov-trajectory}
Suppose the Markov chain $(x_t)_{t\geq 0}$ is aperiodic, and let $\varepsilon \in (0,D)$, $\delta \in (0,1)$.
By observing a trajectory of length $N=\widetilde O\!\left(\frac{D^2\card{\mathcal{X}}^2n}{\varepsilon^2}
       +\tau\frac{D\card{\mathcal{X}}}{\varepsilon}
    \right)$, with probability at least $1-\delta$ we find in polynomial time a policy with average cost at most $g^* + \varepsilon$ per step. Here, $\tau$ denotes the mixing time of the underlying Markov chain $(x_t)_t$, and $\widetilde O$ suppresses logarithmic factors in
$n,\card{\mathcal{X}},D, 1/\varepsilon,1/\delta$, and $1+\tau$.
\end{restatable}

\subsection{Related work}
\label{sec:related-work}

\paragraph{Adversarial MTS.} In the classical adversarial request model, an online algorithm is judged based on its competitive ratio -- worst case ratio of algorithm's cost and the cost of the best solution in hindsight. It is well understood what competitive ratios are achievable for MTS in this model: For any $n$-point metric space, the best competitive ratio of deterministic algorithms is precisely $2n-1$ \cite{MTSintroduction}, while for randomized algorithms it is in $\Omega(\log n) \cap O(\log^2 n)$ (with both the lower and upper bounds being tight for some metrics) \cite{MTSBubeck, MTSCoester, BubeckCR23}. Better bounds are achievable for special cases of MTS: For example, the $k$-server problem on an $n$-point metric is an MTS in an $\binom{n}{k}$-point metric, but admits a deterministic algorithm with competitive ratio $k$ \cite{CoesterKZ26}.

\paragraph{Beyond worst-case analysis.} The adversarial model has been widely criticized as being ``too crude and unrealistic'' \cite{beyondcompanalysis}, because it often produces overly pessimistic guarantees and can obscure meaningful differences between algorithms that behave quite differently in practice. To overcome this criticism, several research directions have emerged.
A prominent one is that of \emph{learning-augmented algorithms}, where online algorithms are given access to (possibly inaccurate) predictions which are meant to expose
common patterns in the data and lead to improved performance. There has been extensive work in this setting on MTS \cite{AntoniadisCEPS23online, Christianson23, AntoniadisCEPS23mixing, CosaE25} and special cases of it, such as $k$-server ~\cite{LindermayrMS25,Christianson23}, caching (also known as paging) \cite{LykourisV21,Rohatgi20,Wei20,AntoniadisCEPS23online,JiangPS22,BansalCKPV22,AntoniadisBEFHLPS23},
convex function chasing~\cite{ChristiansonHW22,LechowiczC0BHWS24}, and dynamic power management~\cite{AntoniadisCEPS21}. Another setting in which MTS has been studied is \emph{smoothed analysis}, where the performance of an algorithm is evaluated on random perturbations of worst-case instances \cite{Sch05, CoesterUmen25}.

Another natural approach, which we pursue in this paper, is to \textbf{model the input as a stochastic process}, rather than assuming it is generated adversarially. IID models have already been studied in the context of various online problems, including 
matching \cite{Gupta19}, Steiner tree/forest \cite{Garg08}, set cover \cite{Grandoni13, Gupta24}, and special cases of MTS such as $k$-server \cite{Stochastic-kserver-Uber, decentralized-kserver} and list update \cite{IIDListUpdate}. Out of all prior work, Markov paging, introduced by \cite{MarkovPaging}, is the closest in spirit to our setting. In their setup, requests are generated by a (known) Markov chain on the page universe, and the goal is to minimize the expected cache miss rate. \cite{MarkovPaging} note that the optimal online policy can be found precisely, but in exponential time,\footnote{Even though a nearly optimal policy can be found in polynomial time for stochastic MTS, this does not imply polynomial time in the natural parameters for paging, as the reduction from paging to MTS incurs an exponential blowup in the state space.} and show that an $O(1)$-approximation of the optimal policy can be found in polynomial time. The best polynomial-time approximation ratio was subsequently improved to 4 \cite{Lund99} and recently to 2 \cite{NewMarkovPaging}.

\paragraph{Markov decision processes and fixed-point methods.}
Average-reward Markov decision processes (MDPs) \cite{Howard60, Blackwell62} are a classical framework for infinite-horizon decision-making under uncertainty. An MDP transitions between multiple states, where the next state is drawn from a probability distribution that depends on the current state and the chosen action. Each state-action pair provides a certain reward, and the goal is to design a policy which chooses actions in order to maximize the average reward per step over an infinite time-horizon. Classical theory \cite{Puterman94, Bert95, SuttoBarto18} characterizes optimal policies through the Bellman optimality equation, which can be solved for example via policy iteration or value iteration.
A key technical difference from discounted MDPs is that the average-reward Bellman operator is non-expansive but not necessarily contracting, and therefore ordinary value iteration need not converge. Halpern iteration \cite{Halpern67}, also known in this context as anchored value iteration, provides a method to overcome this issue by computing approximate fixed points of non-expansive maps. Motivated by reinforcement learning applications, where the environment is not known a priori and must be learned through interaction, a significant line of work, such as \cite{ZhangXie23, ZurekChen24, LeeBravCom25, ZurekChen25}, studies the sample complexity of computing near-optimal policies under a generative model introduced by \cite{KearnsSingh98}, in which transition probabilities are unknown but independent samples of next states can be obtained for any given state-action pairs.

\section{Preliminaries}
\label{sec:preliminaries}

\paragraph{MDP formulation.} MTS in the Markov requests model can be cast as an MDP where each state is represented by a tuple $(s, x, c)$, with $s \in M$, $x \in \mathcal{X}$, and $c \colon M \rightarrow \reals_{\geq 0} \cup \{\infty\}$, and where each available action corresponds to a server location in $M$. Since our model allows infinitely many possible cost functions $c$, the state space of the MDP in this reduction is infinite, and therefore previous algorithms for finding a nearly optimal policy for finite-state MDPs do not apply directly. Our algorithms instead exploit the particular structure of stochastic MTS.

When considering the most general class of online policies, namely \emph{non-stationary} policies, a policy makes
a decision based on everything it has observed so far (i.e., its own previous decisions $s_0, \dots, {s_{t-1}}$, cost functions $c_1, \dots, c_t$, and states $x_0, \dots, x_t$ of the chain generating the requests). Such
a policy can be deterministic, mapping a history to the next
state, or randomized, mapping the history to a probability distribution over states. If a policy makes a decision by only taking into account $s_{t-1}$, $c_t$, and $x_t$, then the policy is called \emph{stationary}. To define the average cost per step for a  policy $\pi$, we follow the definition in the MDP literature \cite{Puterman94}. Let the path induced by $\pi$ be $(a_t)_{t\geq 0}$. Then the cost of $\pi$ is defined as 
\[
g(\pi) \coloneqq \limsup_{T \rightarrow \infty} \frac{1}{T}\E\left[\sum_{t=1}^T d(a_{t-1}, a_t) + c_t(a_t)\right].
\]\

\paragraph{Overview of techniques.} 
Our analysis has three main ingredients. First, we characterize the structure of optimal policies through optimality equations. These equations draw on classical optimality equations \cite{Puterman94}, but in our models they involve averaging the Bellman operator over the randomness in the request process. Second, we follow a standard approach \cite{Puterman94} and show that a small Bellman residual error leads to a nearly-optimal online policy. Third, we show how to obtain an arbitrarily small Bellman residual error by employing Halpern iteration \cite{Halpern67}, a method used for finding fixed points in non-expansive maps. Since the request distribution is unknown, we run the iterations using an empirical mean of the Bellman operator applied on samples, and control the discrepancy between the empirical and true operators by concentration bounds. 

\paragraph{Paper organization.} We first consider the special case of IID requests in section \ref{sec:IID-requests}. In section \ref{sec:Markov-requests}, we show how our results extend to the case of Markov requests. In section \ref{sec:experiments}, we provide a short experimental evaluation of our proposed algorithms. 

\section{IID requests}
\label{sec:IID-requests}
This section is concerned with the case of \emph{IID requests}, where we assume that each service cost function $c_t$ is drawn independently from an unknown distribution $\mathcal{D}$. We assume that $g^*$, the optimal expected cost per step, is finite, which implies that $\mathbb E_{c\sim\mathcal D}[\min_s c(s)]<\infty$.

For a service cost function $c$, the corresponding Bellman operator $B^c\colon \reals^M\to\reals^M$ is defined by\begin{equation} \label{eq:bellman-op}(B^c u)(s)\coloneqq\min_{s'\in M}\{d(s,s')+c(s')+u(s')\}.\end{equation}
For ease of notation, we will henceforth write $B^c u(s)$ to denote $(B^c u)(s)$.

For $u\colon M\rightarrow \reals$, define the stationary deterministic policy $\pi_u$ by 
\begin{equation}
\label{eq:define-policy}
\pi_u(s, c) \coloneqq \argmin_{s' \in M} \{d(s,s') + c(s') + u(s') \}. 
\end{equation}

That is, when the current state is $s\in M$ and the service cost function $c$ is received, the policy $\pi_u$ moves to $\pi_u(s, c)$. Intuitively, we think of $u(s)$ as an approximation for the cost to be paid to serve the remaining requests if starting from $s$. Note that shifting $u$ by an additive constant still induces the same policy.

We prove that if $u^*$ satisfies the property below, then the policy $\pi_{u^*}$ is optimal.

\begin{claim}[Structure of optimal policy] 
\label{claim:structure-of-opt}
Suppose $u^*\colon M \rightarrow \reals$ and $g^* \in \reals$ satisfy
\begin{equation}
    \label{eq:opt}
 \E_{c \sim \mathcal{D}} [B^c u^*(s)] - u^*(s) = g^* \quad \text{for all } s \in M.
\end{equation}

Then the policy $\pi^* \coloneqq \pi_{u^*}$ is an optimal policy, even among non-stationary randomized policies, and achieves average cost $g(\pi^*) = g^*$ per step.
\end{claim}
\begin{proof}
    Let $a_0, a_1, \dots, a_T$ be random variables describing the path taken by some non-stationary, possibly randomized policy $\pi$ for the first $T$ time steps. Here, $a_t$ depends on $a_0, \dots, a_{t-1}$ and $c_1, \dots, c_t$. The cost of this policy up to time $T$ is 
    \[
    \E \left[\sum_{t=1}^T d(a_{t-1}, a_t) + c_t(a_t)\right].
    \]
    By the definition of $\pi^*$, and by denoting $o_t \coloneqq \pi^*(a_{t-1}, c_t)$ (i.e. what the policy $\pi^*$ would have done at time $t$ if starting from $a_{t-1}$), we get
    \begin{equation}    
    \label{eq:subopt-move}
    d(a_{t-1}, a_t) + c_t(a_t) \geq  d(a_{t-1}, o_t) + c_t(o_t) + u^*(o_t) - u^*(a_t).
    \end{equation}
    
    Plugging this in, we get that the expected cost of policy $\pi$ for the first $T$ time steps is at least 
    \[
    \E \left[\sum_{t=1}^T d(a_{t-1}, o_t) + c_t(o_t) + u^*(o_t) - u^*(a_t)\right].
    \]
    By the definition of $o_t$ and $\pi^*$, this is equal to 
    \[
    \E \left[\sum_{t=1}^T B^{c_t}u^*(a_{t-1}) - u^*(a_t)\right].
    \]
    
    Since $\E[B^{c_t}u^*(a_{t-1}) \mid a_{t-1}] = u^*(a_{t-1}) + g^*$ by \eqref{eq:opt}, we get that the expression above is equal to
    \begin{align*}
    \sum_{t=1}^T \E \left[\E \left[ B^{c_t}u^*(a_{t-1}) \mid a_{t-1}\right]\right] - \E [u^*(a_t)] &= 
    \sum_{t=1}^T \E \left[ u^*(a_{t-1}) + g^* - u^*(a_t)\right] \\
    &= \E [u^*(a_0) - u^*(a_T)] + g^*T.
    \end{align*}
        
    By dividing by $T$ and letting $T \rightarrow \infty$, the first term vanishes and we obtain $g(\pi) \geq g^*$.

    Repeating essentially the same calculations with $a_t$ defined recursively by $a_t \coloneqq \pi^*(a_{t-1}, c_t)$, we get that the average cost per step for policy $\pi^*$ is $g(\pi^*) = g^*$. This is because $o_t = a_t$, and therefore the only inequality involved, namely \eqref{eq:subopt-move}, turns into equality.
\end{proof}

By defining the operator $\bar{B}$ such that $\bar{B} u \coloneqq  \E_{c \sim \mathcal{D}} [B^c u]$,
\eqref{eq:opt} can be interpreted as $u^*$ being a fixed point of $\bar{B}$, up to an additive constant $g^*$. We next show the existence of such a pair $(u^*, g^*)$.

\begin{claim}[Existence of an optimal policy]
    There exist $u^*$ and $g^*$ satisfying \eqref{eq:opt}, with $\| u^*\|_\infty \leq D$.
\end{claim}
\begin{proof}
Let $H \coloneqq \{u\colon M \rightarrow [0,D]\}$, and let $T\colon H \rightarrow H$ be defined by $T u \coloneqq \bar{B} u - \min(\bar{B}u)$. Note that $Tu \in H$ for any $u$ because $B^c u$ is 1-Lipschitz with respect to the underlying metric space, which implies that $Tu$ is 1-Lipschitz as well. Since $H$ is convex and compact, and $T$ is continuous, by Brouwer's fixed-point theorem we know that $T$ has a fixed point $u^* \in H$. By setting $g^* \coloneqq \min(\bar{B} u^*)$, we get $\bar{B} u^* \equiv u^* + g^*$, as required in \eqref{eq:opt}.
\end{proof}

Note that $g^*$ is unique, but $u^*$ is not necessarily unique. Next, we bound the error of a policy $\pi_u$.

\begin{claim}[Bounding the policy error]
\label{cl:bound-policy-error}
Let $\pi_u$ be the policy induced by $u$, and let $g(\pi_u)$ be its cost. Then
    \[g(\pi_u) - g^* \leq  \|\bar{B} u - u\|_{\SP},\]
where $\|\cdot\|_\SP$ denotes the span seminorm, 
defined by \[ \|u\|_{\SP}\coloneqq\max(u) - \min(u).\footnote{Throughout the paper, we use $\max(u)$ and $\min(u)$ as a shorthand for $\max_{s \in M}u(s)$ and $\min_{s \in M}u(s)$, respectively.}\] 
\end{claim}
\begin{proof}
Let $a$ be the trajectory induced by $\pi_u$, recursively defined by $a_t \coloneqq \pi_u(a_{t-1}, c_t)$. The expected cost of $\pi_u$ for the first $T$ time steps is 
    \begin{align*}
    \E \left[\sum_{t=1}^T d(a_{t-1}, a_t) + c_t(a_t)\right] &= \E\left[ \sum_{t=1}^T d(a_{t-1}, a_t) + c_t(a_t) + u(a_t) - u(a_t)\right] \\
    &= \E\left[ \sum_{t=1}^T B^{c_t}u(a_{t-1}) - u(a_t)\right] \\
    &= \E\left[ \sum_{t=1}^T B^{c_t}u(a_{t-1}) - u(a_{t-1})\right] + \E[u(a_0) - u(a_T)] \\ 
    &= \E\left[ \sum_{t=1}^T \bar{B}u(a_{t-1}) - u(a_{t-1})\right] + \E[u(a_0) - u(a_T)] \\
    &\leq T \max(\bar{B}u - u) + \E[u(a_0) - u(a_T)].
    \end{align*}

Here, in the second equality we used the definitions of $B^{c_t}u$ and $a_t$, and in the penultimate step we used the definition of $\bar{B}$. Dividing by $T$ and letting $T\rightarrow \infty$, we  obtain 
\begin{equation}
\label{eq:bounding-g-piu}
g(\pi_u) \leq \max(\bar{B}u - u).
\end{equation}

 Let $o$ be the trajectory induced by $\pi^*$, recursively defined by $o_t \coloneqq \pi^*(o_{t-1}, c_t)$. Similarly to the above, we show that the expected cost of policy $\pi^*$ for the first $T$ time steps is 
 \begin{align*}
    \E \left[\sum_{t=1}^T d(o_{t-1}, o_t) + c_t(o_t)\right] &= \E\left[ \sum_{t=1}^T d(o_{t-1}, o_t) + c_t(o_t) + u(o_t) - u(o_t)\right] \\
    &\geq \E\left[ \sum_{t=1}^T B^{c_t}u(o_{t-1}) - u(o_t)\right] \\
     &= \E\left[ \sum_{t=1}^T B^{c_t}u(o_{t-1}) - u(o_{t-1}) \right] + \E[u(o_0) - u(o_T)] \\
      &= \E\left[ \sum_{t=1}^T \bar{B}u(o_{t-1}) - u(o_{t-1})\right] + \E[u(o_0) - u(o_T)] \\
      &\geq T \min(\bar{B}u - u) + \E[u(o_0) - u(o_T)].
    \end{align*}

Dividing by $T$ and letting $T\rightarrow \infty$, we  obtain 
\begin{equation}
\label{eq:bounding-g*}
g^* = g(\pi^*) \geq \min(\bar{B}u - u).
\end{equation}

Putting \eqref{eq:bounding-g-piu} and \eqref{eq:bounding-g*} together, we obtain 
\[
\min(\bar{B}u - u) \leq g^* \leq g(\pi_u) \leq \max(\bar{B}u - u),
\]
from which we conclude that 
\[
g(\pi_u) - g^* \leq \|\bar{B}u - u\|_{\SP}.\qedhere
\]
\end{proof}

Equipped with Claim \ref{cl:bound-policy-error}, we now seek to find $u$ for which $\|\bar{B}u - u\|_{\SP} \leq \varepsilon$. To this end, we make use of Halpern iteration and its convergence guarantees. 

\begin{restatable}{theorem}{Halpernbound}{(Halpern residual bound).}\label{thm:halpern-rate2}
Let $F\colon Z \rightarrow Z$ be a non-expansive map defined on space $Z$ with associated seminorm $\|\cdot \|$. Let $C \subseteq Z$ be a convex set, and let $\diam(C) \coloneqq \sup_{z, z' \in C} \| z-z'\|$. Let $(\alpha_k)_{k\ge1}$ be a step size sequence with $\alpha_k \in (0,1)$. The Halpern iteration \cite{Halpern67} with anchor at $0$ and starting point $z_0 \coloneqq 0$ produces a sequence $(z_k)_{k\ge0}$ defined by
\begin{equation}\label{eq:halpern2}
z_{k} \coloneqq (1-\alpha_{k}) Fz_{k-1},\qquad k\geq 1.
\end{equation}
Suppose that $0\in C$, and $z \in C$ implies $Fz \in C$. Then, for the choice
\[
\alpha_k=\frac{2}{k+2},\qquad k\ge 1,
\]
we have 
\[
\|z_k-Fz_k\| \leq \frac{4 \cdot \diam(C)}{k+1}.
\]
\end{restatable}

To the best of our knowledge, \cite{SaSh17} were the first to prove this bound for general normed spaces, as previous work had focused on Hilbert spaces and smooth Banach spaces. For completeness, we include a proof of Theorem \ref{thm:halpern-rate2} in Appendix \ref{sec:halpern-bound-proof}.

In order to apply Halpern iteration, we show that the Bellman operator is non-expansive.

\begin{claim} \label{cl:non-expansive}
    $B^c$ is non-expansive with respect to $\| \cdot \|_\SP$, for any $c\colon M \rightarrow \reals_{\geq 0} \cup \{\infty\}$ with $\min(c)<\infty$.
\end{claim}
\begin{proof}
    Let $u, v \in \reals^M$. Fix $s \in M$ and let $r \in M$ such that $B^c u(s) = d(s,r) + c(r) + u(r)$.
    
    Then 
    $B^c v(s) \leq d(s,r) + c(r) + v(r)$, and therefore \[B^c v(s) - B^c u (s) \leq v(r) - u(r) \leq \max(v-u).\]
    Similarly, \[B^c v(s) - B^c u (s) \geq \min(v-u),\]
    and we can conclude that 
    \[
    \|B^c v- B^c u \|_\SP \leq \| v-u \|_\SP. \qedhere
    \]
\end{proof}

If $\mathcal{D}$ were known, and if we could compute $\bar{B} u$ for any $u$, then we could find an approximately optimal online policy by running Halpern iteration for $\bar{B}$. But since we assume a weaker, generative model, we do not know $\mathcal{D}$. Instead, we sample $N$ cost functions $c_1, \dots, c_N$ from $\mathcal{D}$, where $N$ will be determined later. We then define an empirical counterpart to $\bar{B}$, by
\[
\widetilde{B}u \coloneqq \frac{1}{N} \sum_{i=1}^N B^{c_i}u.
\]

We determine how many samples $N$ are necessary to ensure that $\widetilde{B}$ is a good approximation of $\bar{B}$. 

\begin{restatable}{claim}{samplecomplexityiid}{(Sample complexity).}
\label{cl:sample-complexity}
Let
$
C \coloneqq \{u \in \mathbb R^M \colon \|u\|_\SP \le D\}.
$
Let $\delta \in (0,1)$ and $\varepsilon' > 0$. Suppose that
%$c_1,\ldots,c_N$ are drawn independently from $\mathcal D$ and
\[
N \ge
\frac{8D^2}{\varepsilon'^2}
\left(
    n \log\left(\left\lceil \frac{16D}{\varepsilon'} \right\rceil + 1\right)
    +
    \log \frac{2n}{\delta}
\right).
\]
Then, with probability at least $1-\delta$ it holds simultaneously for all
$u \in C$ that
$
\|\widetilde B u - \bar B u\|_\SP \le \varepsilon' .
$
\end{restatable}

Claim \ref{cl:sample-complexity} follows from Claim \ref{cl:uniform-estimation-oracles}, proved in Appendix \ref{sec:omitted-proofs} for the more general setting of Markov requests. 

Putting everything together, we are now ready to prove Theorem \ref{th:culminating-iid}.
\begin{proof}[Proof of Theorem \ref{th:culminating-iid}]
We apply Theorem \ref{thm:halpern-rate2} using $\widetilde B$. Note that
$\widetilde B$ is non-expansive with respect to $\|\cdot\|_\SP$,
because it is an average of the maps $B^{c_1},\ldots,B^{c_N}$, which are
non-expansive by Claim \ref{cl:non-expansive}.

Let
$
C \coloneqq \{u \in \mathbb R^M \colon \|u\|_\SP \le D\}.
$
Since $B^c u$ is $1$-Lipschitz with respect to the underlying metric, we have
$\|B^c u\|_\SP \le D$ for every $u$, so $\widetilde B$ maps
$C$ to itself. Moreover, $\diam(C) = 2D$.

Run the Halpern iteration for
$
k \coloneqq \left\lceil \frac{16D}{\varepsilon} \right\rceil
$ steps. By Theorem \ref{thm:halpern-rate2}, the empirical residual error satisfies
\begin{equation}
\|z_k - \widetilde B z_k\|_\SP
\le
\frac{4 \cdot \diam(C)}{k+1}
\le
\frac{\varepsilon}{2}.
\label{eq:bounding-error-Btilde}
\end{equation}

Combining \eqref{eq:bounding-error-Btilde} with Claim \ref{cl:sample-complexity} applied with $\varepsilon' = \varepsilon/2$, and by using the triangle inequality, we get that
    \[\|z_k - \bar{B} z_k\|_\SP \leq \varepsilon\]
holds with probability at least $1-\delta$. Finally, by Claim \ref{cl:bound-policy-error}, we obtain that the policy $\pi_{z_k}$ induced by $z_k$ satisfies $g(\pi_{z_k}) \leq g^* + \varepsilon$ with probability at least $1-\delta$.
\end{proof}

\section{Markov requests}
\label{sec:Markov-requests}

We consider requests generated by a Markov chain $(x_t)_{t\geq0}$ on a known finite
state space $\mathcal X$.
Conditional on $x_{t-1}=x$, the pair $(x_t,c_t)$ is drawn from
$\mathcal D_x$, independently of the preceding history.
The chain is irreducible, with stationary distribution $p$, and the
online algorithm observes both $x_t$ and $c_t$ before choosing its move.
The distributions ($\mathcal D_x)_{x\in\mathcal{X}}$ and $p$ are unknown.
We assume that $g^*$, the optimal expected cost per step, is finite. By irreducibility, this implies that
$\mathbb E_{(x',c)\sim\mathcal D_x}[\min_{s\in M}c(s)]<\infty$
for every $x\in\mathcal X$, ensuring that the Bellman expectations
below are finite. 

%\paragraph{Optimal policies and residual error.}
Instead of identifying a stationary policy by a function $u\colon M \rightarrow \reals$ as in the IID case, we now need to consider a separate function $u_x\colon M \rightarrow \reals$ for each $x \in \mathcal{X}$. We think of $u_x(s)$ as an approximation for the relative cost to be paid to serve the remaining requests if the server is located at $s\in M$ and the chain is in state $x \in \mathcal{X}$. Each such $u$ induces a stationary deterministic policy $\pi_u$, defined by 
\begin{equation}
\label{eq:define-policy-Markov}
\pi_u(s, c, x) \coloneqq \argmin_{s' \in M} \{d(s,s') + c(s') + u_x(s') \}. 
\end{equation}
That is, when the MTS state is $s\in M$, the next revealed cost function is $c$, and the next revealed Markov chain state is $x$, the policy $\pi_u$ moves to $\pi_u(s, c, x)$.
Note that adding a constant to any $u_x$ does not change the induced policy.

We define \[(\overline Bu)_x\coloneqq\mathbb E_{(x',c)\sim\mathcal D_x}[B^cu_{x'}],\]
where $B^c$ is defined in~\eqref{eq:bellman-op}. The following claims extend the corresponding results of
Section~\ref{sec:IID-requests}, and their proofs can be found in Appendix ~\ref{app:markov-structure}.

\begin{restatable}{claim}{optpolicyMarkov}{(Structure of an optimal policy).}
\label{claim:structure-of-opt-Markov}
Suppose $u^*\in(\mathbb R^M)^{\mathcal X}$ and
$\lambda^*\in\mathbb R^{\mathcal X}$ satisfy
\begin{equation}
    (\overline Bu^*)_x(s)-u_x^*(s)=\lambda_x^*
    \qquad\text{for every }x\in\mathcal X, s\in M.
    \label{eq:opt-Markov}
\end{equation}
Then $\pi_{u^*}$ is optimal, even among non-stationary randomized
policies, and $g^*\coloneqq g(\pi_{u^*})=\sum_xp_x\lambda_x^*$.
\end{restatable}

\begin{restatable}{claim}{existenceMarkov}{(Existence of an optimal policy).}
\label{cl:existence-Markov}
There exist $u^*$ and $\lambda^*$ satisfying~\eqref{eq:opt-Markov}.
\end{restatable}

For all three observation
models that we consider, we denote the observations by $(x_i,c_i,x_i')_{i=1}^N$.
The way in which $x_i$ is obtained differs between models: with conditional oracles, $x_i$ can be chosen by the learner; with a stationary oracle, $x_i$ is drawn from the stationary distribution $p$ independently from previous history; in the observed trajectory model, $x_{i+1}=x'_i$. In every model, conditional on $x_i$ and the preceding observations,
the pair $(x_i',c_i)$ is drawn from $\mathcal D_{x_i}$.

Let $I_x\coloneqq\{(x'_i,c_i):x_i=x\}$ be the multiset of observations with source state $x$, and let $N_x\coloneqq\vert I_x\vert$. Define
\begin{equation}
    (\widetilde Bu)_x\coloneqq
    \begin{cases}
        \displaystyle\frac1{N_x}\sum_{(x',c)\in I_x}B^{c}u_{x'},&N_x>0,\\[1ex]
        0,&N_x=0.
    \end{cases}
    \label{eq:trajectory-empirical-operator}
\end{equation}

\begin{restatable}{claim}{nonexpansiveMarkov}{(Non-expansiveness).}
\label{cl:nonexpansive-Markov}
$\widetilde B$ is non-expansive under
$\|\cdot\|$, where $\|v\| \coloneqq \max_{x\in \mathcal{X}}\|v_x\|_{\SP}$.
\end{restatable}

The following bound weights estimation errors by the stationary
probabilities, allowing states that occur less frequently to be
estimated less accurately.

\begin{restatable}{claim}{boundingpolicyerrorMarkov}{(Bounding the policy error).}
\label{cl:bound-policy-error-Markov}
For every $u\in(\mathbb R^M)^{\mathcal X}$, conditional on observations $(I_x)_{x\in\mathcal{X}}$,
\begin{equation}
\begin{aligned}
    g(\pi_u)-g^*
    \le \sum_{x\in \mathcal{X}} p_x\|(\overline Bu)_x-u_x\|_{\SP}
    \le \|\widetilde Bu-u\|
       +\sum_{x\in \mathcal{X}} p_x\|(\overline Bu)_x-(\widetilde Bu)_x\|_{\SP}.
\end{aligned}
\label{eq:markov-weighted-policy-error}
\end{equation}
\end{restatable}

\begin{lemma}\label{lemma:to-conclude-Markov} Let $C\coloneqq\{u\in(\mathbb R^M)^{\mathcal X}\colon\|u_x\|_{\SP}\le D
\text{ for all }x \in \mathcal{X}\}$.
If \[\sum_xp_x \|(\overline Bu)_x-(\widetilde{B}u)_x\|_{\SP}\leq\varepsilon/2 \text{ for all } u\in C,\] the policy $\pi_{z_k}$ obtained after running $k=O(D/\varepsilon)$ Halpern iterations has cost at most $g^*+\varepsilon$. 
\end{lemma}
\begin{proof}
Since $\|B^c u_{x'}\|_{\SP} \le D$ for all $(x',c)\in I_x$, $\widetilde B$ maps $C$ to itself. To bound $\|(\widetilde Bu)-u\|$, we run Halpern iteration~\eqref{eq:halpern2} using
$\widetilde B$ and set
$u\coloneqq{z_k}$ for $k\coloneqq\lceil16D/\varepsilon\rceil$. Since $C$ is convex, contains $0$, and has diameter at most
$2D$ with respect to $\|\cdot\|$, Theorem~\ref{thm:halpern-rate2} and Claim~\ref{cl:nonexpansive-Markov}
give $z_k\in C$ and
\begin{equation*}
    \|\widetilde Bz_k-z_k\|
       \le \frac{8D}{k+1}\le\frac{\varepsilon}{2}.
    \label{eq:trajectory-empirical-residual}
\end{equation*}
Combining our hypothesis applied to $u=z_k$ with Claim \ref{cl:bound-policy-error-Markov} concludes the proof.
\end{proof}

To complete the proofs of Theorems \ref{th:culminating-Markov} and \ref{th:culminating-Markov-trajectory}, it suffices to show that, with probability at least $1-\delta$, we have $\sum_xp_x
          \|(\overline Bu)_x-(\widetilde{B}u)_x\|_{\SP} \leq \varepsilon/2$ for all $u\in C$. Since the bound holds simultaneously for all $u\in C$, it applies
to $z_k$ even though this iterate depends on the observations $(x_i, c_i, x'_i)_{i=1}^N$. We prove the bound in
Appendix ~\ref{app:concentration}. 

\section{Experiments}
\label{sec:experiments}
\subsection{Synthetic caching}

We evaluate our algorithm on an important special case of MTS, namely the caching problem. In this problem, there are $n$ pages and $k$ cache slots. Each request consists of a page. If the requested page is not in the cache, a cache miss occurs and unit cost must be paid. In this case, the page must be brought into the cache, and a different page might need to be evicted to free up space. 

We conduct a synthetic experiment in which the requests are generated according to a Markov chain on the pages. The pages are arranged in a circle, and the probability that $j$ is requested next, conditioned on the current request being $i$, decays exponentially with $i \cdot \dist(i, j)$.
This construction yields a family of conditional distributions of the next request ranging from uniform (for $i=0$) to highly concentrated (for large $i$). We compare our algorithm against: 1) the Dominating Distribution Algorithm (DDA), which is guaranteed to be within a factor of $2$ of the optimal online policy \cite{NewMarkovPaging}; 2) the Work Function Algorithm (WFA), (one of) the best possible classical online deterministic algorithms, obtaining a competitive ratio of $k$ \cite{CoesterKZ26}; 3) the Least Recently Used (LRU) Algorithm, which upon a cache miss evicts the page which was requested least recently, also achieving a competitive ratio of $k$ \cite{CompetitiveRatioDef} and often obtaining a very good performance in practice; 4) the Randomized Marker Algorithm, which achieves the optimal competitive ratio of $O(\log k)$ among randomized algorithms \cite{pagingfiat}. DDA is given access to the transition matrix of the underlying Markov chain, while our algorithm operates without such knowledge and instead learns from samples. We set the number of pages to $n=12$ and vary the size $k$ of the cache. We run $20$ tests for each value of $k$, each test consisting of \(10{,}000\) requests. In each test, our algorithm uses $500$ samples for each Markov state and runs Halpern iteration for $100$ steps. 

\begin{figure*}[t] \centering \begin{subfigure}[t]{0.53\textwidth} \centering \includegraphics[width=\linewidth]{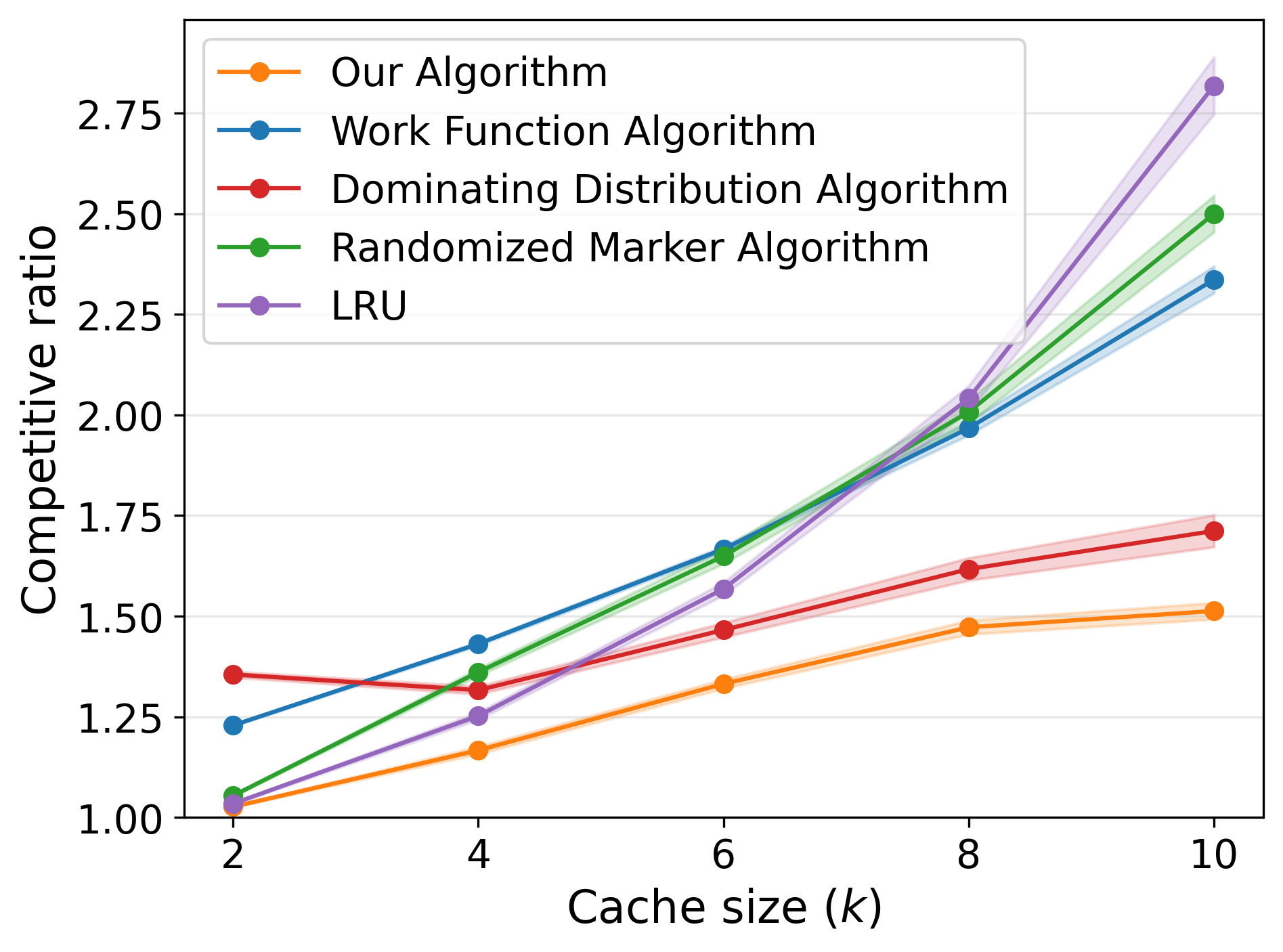} \caption{Experiment 1: Markov caching. Comparison of the competitive ratios of algorithms, as the cache size varies.} \label{fig:plot1} \end{subfigure} \hfill \begin{subfigure}[t]{0.44\textwidth} \centering \includegraphics[width=0.95\linewidth]{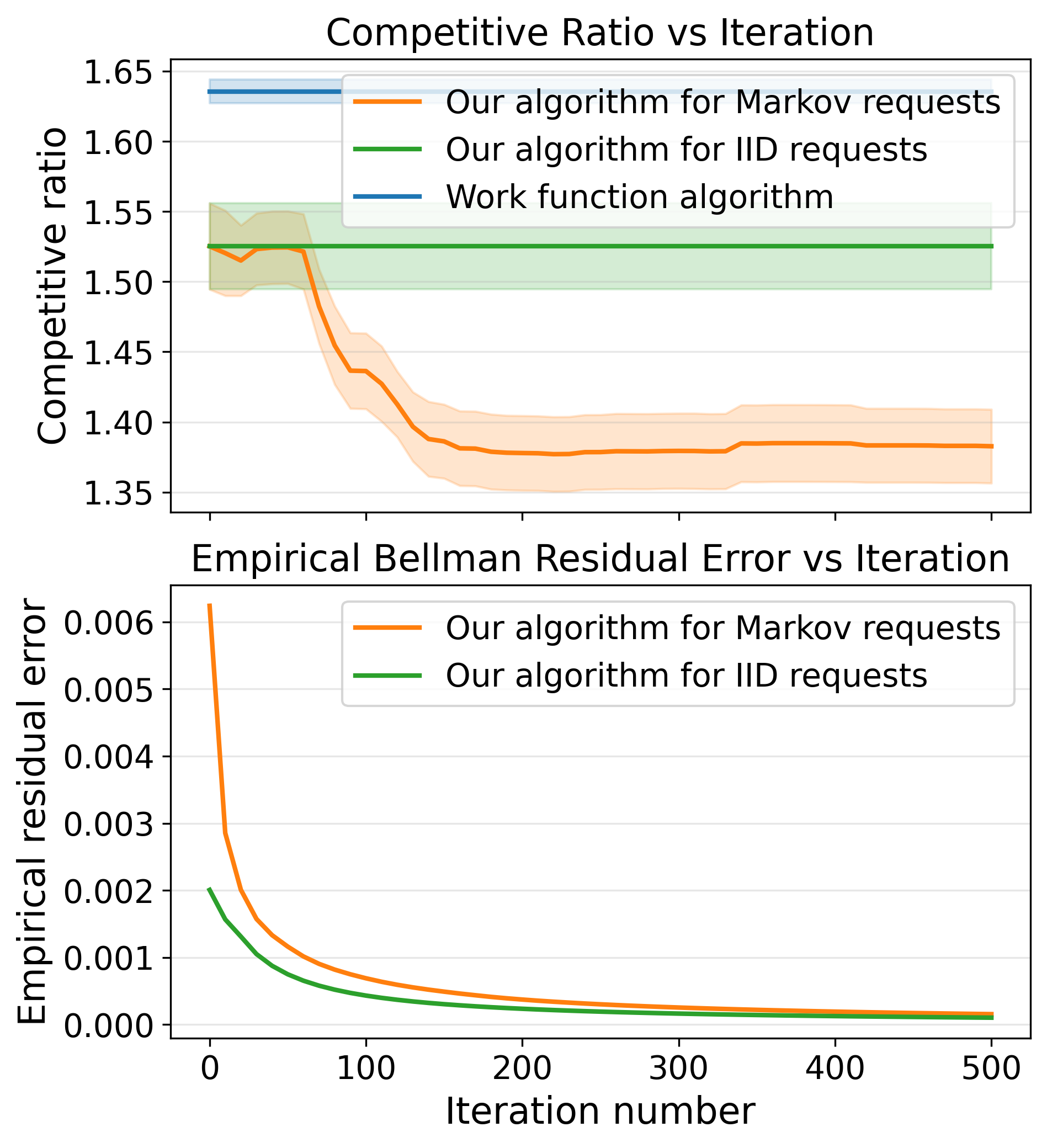} \caption{Experiment 2: citibike $k$-server. The competitive ratios and empirical Bellman residual errors, as a function of the number of Halpern iterations.} \label{fig:plot2} \end{subfigure} \caption{Results of our experiments. Shades denote standard error of the mean, computed using numpy.std() and dividing by the square root of the number of runs.} \label{fig:2plots} \end{figure*}

Fig. \ref{fig:plot1} shows the empirical competitive ratios obtained by the algorithms. The competitive ratio is defined as the ratio between the cost of an online algorithm and the cost of the optimal offline algorithm.\footnote{We use the optimal offline algorithm as a benchmark in all of our experiments because the experiments with real-world data are not generated by an actual Markov chain, so there is no optimal online policy to compare against.} We compute the latter quantity by dynamic programming. We observe that our algorithm consistently has the best performance. The factors of variability in this experiment are the randomly generated problem instance and the internal randomness used by DDA and the Randomized Marker Algorithm.

\subsection{$k$-server on a real-world dataset}
We additionally evaluate our algorithms on a real-world dataset from a popular bike-sharing platform {citibike} \cite{citibike_system_data}. We consider the $k$-server problem \cite{Manasse88}, a central problem in online algorithms and a special case of MTS: There are $k$ servers which must be moved in a metric space of size $n$ to serve incoming requests. Each request is a point in the metric space, and must be served by moving one of the servers there. The goal is to minimize the total distance traveled by the servers. We consider the $k$-server problem with $k=n-2$, which implies that at any moment there are precisely $2$ locations not covered by the servers. This is also known as the $2$-evader problem \cite{2evader}.

We use the citibike dataset as follows: We set $n=10$ and take the top $10$ most popular starting stations with Euclidean distance to be our metric space. We then map each trip starting in one of these stations to a request. We model temporal patterns in the requests by using a Markov chain with state space $\{(w, h) \mid w\in \{0,1\}, h\in \{0, \dots, 23\}\}$, where $w$ indicates whether the request occurred on a weekday or weekend, and $h$ indicates the hour of the day. Here, $w$ and $h$ are observed context labels used to construct an approximate Markov model. We run our algorithm for Markov requests (where we create for each request an observed Markov state as described above), our algorithm for IID requests, and the Work Function Algorithm (WFA). The latter obtains the optimal competitive ratio of $n-2$ among classical deterministic online algorithms for the $2$-evader problem \cite{2evader}. We create $12$ instances, corresponding to the months of $2025$, and evaluate the performance of our algorithms using leave-one-out cross validation. We run $500$ Halpern iterations and evaluate our policies after every $10$ iterations. 

The results are shown in Fig. \ref{fig:plot2}. We observe that our algorithm which assumes IID requests already outperforms WFA. Our algorithm for Markov requests achieves further improvements, despite relying on a relatively simple Markov chain model. We also report the empirical residual error of the Halpern iterates, namely $\|\widetilde{B}z_k - z_k\|$, which approaches $0$ as we find an approximate fixed point of $\widetilde{B}$. This experiment suggests that algorithms designed for the Markov setting (which is often an imperfect model of reality) perform well beyond this idealized abstraction.

The experiments were run on a personal computer with 32 GB RAM and completed in a few hours.

\section{Limitations and conclusion}\label{sec:limitations-and-conclusions}

We develop a general approach for efficiently learning nearly optimal online policies for MTS under IID and Markov requests. This includes applications such as systems with
metric switching costs between operating modes, problems involving
the location of a single movable object (i.e. process migration), and dynamic power management.

However, our guarantees do not yield sample complexity
or running time polynomial in the natural parameters of every
special case of MTS. For example, in the reduction from $k$-server, each MTS state represents an entire
server configuration, so the number of states grows exponentially
in $k$ and can make our approach inefficient. This limitation parallels competitive analysis, where
exploiting the structure of particular problems can give substantially
stronger guarantees than a direct reduction to general MTS.

Perhaps the most fascinating open problem motivated by our work is whether sample complexities polynomial in natural problem parameters of MTS subproblems with special structure can be obtained. A concrete question is whether $k$-server admits a polynomial time algorithm for computing an $O(1)$-approximation of the optimal policy under Markov requests.

\section*{Acknowledgments and disclosures}
\label{sec:acknowledgements}

ChatGPT 6 Astra and predecessor models were used to assist with standard concentration and estimation arguments in Appendix \ref{app:concentration}, as well as with sentence-level rephrasing. The authors independently checked all AI-assisted material and substantially revised it before inclusion in the manuscript. All main ideas in the paper are due to the authors.

Christian Coester is funded by the European Union (ERC, CCOO, 101165139). Views and opinions expressed are however those of the authors only and do not necessarily reflect those of the European Union or the European Research Council. Neither the European Union nor the granting authority can be held responsible for them.

\newpage
\bibliographystyle{alpha}
\bibliography{ref}

\newpage
\appendix

\section{Omitted proofs from section \ref{sec:Markov-requests}} \label{sec:omitted-proofs}

\subsection{Structural properties and policy error bounds}
\label{app:markov-structure}

\optpolicyMarkov*
\begin{proof}
    The proof is similar to the proof of Claim \ref{claim:structure-of-opt}.  Let $a_0, a_1, \dots, a_T$ be random variables describing the path taken by some non-stationary, possibly randomized policy $\pi$ for the first $T$ time steps. Here, $a_t$ depends on $a_0, \dots, a_{t-1}$, $c_1, \dots, c_t$, and $x_0, \dots, x_t$. The cost of this policy up to time $T$ is 
    \[
    \E \left[\sum_{t=1}^T d(a_{t-1}, a_t) + c_t(a_t)\right].
    \]
    By the definition of $\pi_{u^*}$, and by denoting $o_t = \pi_{u^*}(a_{t-1}, c_t, x_t)$ (i.e. what the policy $\pi_{u^*}$ would have done at time $t$ if starting from $a_{t-1}$), we get
    \begin{equation}    
    \label{eq:subopt-move-Markov}
    d(a_{t-1}, a_t) + c_t(a_t) \geq  d(a_{t-1}, o_t) + c_t(o_t) + u^*_{x_t}(o_t) - u^*_{x_t}(a_t).
    \end{equation}
    
    Plugging this in, we get that the expected cost of policy $\pi$ for the first $T$ time steps is at least 
    \[
    \E \left[\sum_{t=1}^T d(a_{t-1}, o_t) + c_t(o_t) + u^*_{x_t}(o_t) - u^*_{x_t}(a_t)\right].
    \]
    By the definition of $o_t$, this is equal to 
    \[
    \E \left[\sum_{t=1}^T B^{c_t}u^*_{x_t}(a_{t-1}) - u^*_{x_t}(a_t)\right].
    \]
    
    Since $\E[B^{c_t}u^*_{x_t}(a_{t-1}) \mid a_{t-1}, x_{t-1}] = u^*_{x_{t-1}}(a_{t-1}) + \lambda^*_{x_{t-1}}$ by \eqref{eq:opt-Markov}, we get that the expression above is equal to
    \begin{align*}
    \sum_{t=1}^T \E \left[\E \left[ B^{c_t}u^*_{x_t}(a_{t-1})\mid a_{t-1}, x_{t-1}\right]\right] - \E [u^*_{x_t}(a_t)] &= 
    \sum_{t=1}^T \E \left[ u^*_{x_{t-1}}(a_{t-1}) + \lambda^*_{x_{t-1}}\right] - \E\left[u^*_{x_t}(a_t)\right] \\
    &= \E [u^*_{x_0}(a_0) - u^*_{x_T}(a_T)] + \E\left[\sum_{t=0}^{T-1} \lambda^*_{x_t}\right].
    \end{align*}
        
    Note that the first term in the expression above is bounded. Dividing by $T$ and letting $T \rightarrow \infty$, the first term vanishes and the second term converges to $\sum_{x} p_x \lambda^*_x$. Therefore, we obtain $g(\pi) \geq \sum_{x } p_x \lambda^*_x$.

    Repeating essentially the same calculations with $a_t$ defined recursively by $a_t \coloneqq \pi_{u^*}(a_{t-1}, c_t, x_t)$, we get that the average cost per step for policy $ \pi_{u^*}$ is $g(\pi_{u^*}) = \sum_{x \in \mathcal{X}} p_x \lambda^*_x$. This is because $o_t = a_t$, and hence inequality \eqref{eq:subopt-move-Markov} turns into equality, and all other equations were already equalities.
\end{proof}

\existenceMarkov*
\begin{proof}
Let $H \coloneqq ([0,D]^M)^\mathcal{X}$, and define the auxiliary operator $T:H\rightarrow H$ by 
\[
(T u)_x \coloneqq (\bar{B} u)_x - \min((\bar{B} u)_x).
\]
Note that $T$ is a normalized version of $\bar{B}$, where normalization takes place along each component $x \in \mathcal{X}$ separately. Since $H$ is convex and compact, and $T$ is continuous, by Brouwer's fixed-point theorem we know that $T$ has a fixed point $u^* \in H$. Setting $\lambda^*_x \coloneqq \min((\bar{B} u^*)_x)$ completes the proof.
\end{proof}

\nonexpansiveMarkov*
\begin{proof}
    Let $u, v \in (\reals^M)^{\mathcal{X}}$. For $x\in\mathcal{X}$ with $N_x>0$, by using the definition of $\widetilde{B}$ and applying triangle inequality,
    \begin{align*}
         \|(\widetilde{B}u)_x - (\widetilde{B}v)_x \|_\SP 
        &= \frac{1}{N_x}\left\| 
         \sum_{(x',c) \in I_x} (B^{c}u_{x'} - B^{c}v_{x'}) \right\|_\SP \\
        &\leq \frac{1}{N_x} \sum_{(x',c) \in I_x} \| 
         B^{c}u_{x'} - B^{c}v_{x'} \|_\SP.
    \end{align*}
    
    Since $B^c$ is non-expansive by Claim \ref{cl:non-expansive}, each term is at most $\|u_{x'} - v_{x'} \|_\SP\leq \max_{z \in \mathcal{X}} \| u_{z} - v_{z} \|_\SP$, so 
    \begin{align*}
         \|(\widetilde{B}u)_x - (\widetilde{B}v)_x \|_\SP 
        &\leq \max_{z \in \mathcal{X}} \| u_{z} - v_{z} \|_\SP.
    \end{align*}
    
    For $x\in\mathcal{X}$ with $N_x = 0$, the same inequality holds trivially because $(\widetilde{B}u)_x = (\widetilde{B}v)_x=0$. Therefore,  
    \[
    \|\widetilde{B}u - \widetilde{B}v \| = \max_{x\in\mathcal{X}} \|(\widetilde{B}u)_x - (\widetilde{B}v)_x \|_\SP \leq \max_{z \in \mathcal{X}} \| u_{z} - v_{z} \|_\SP = \| u - v \|. \qedhere
    \]
\end{proof}

\boundingpolicyerrorMarkov*
\begin{proof}
For ease of notation, let $v_x \coloneqq (\bar{B} u)_x - u_x$. Let $a$ be the trajectory induced by $\pi_u$, recursively defined by $a_t \coloneqq \pi_u(a_{t-1}, c_t, x_t)$. The expected cost of $\pi_u$ for the first $T$ time steps is 
    \begin{align*}
    \E \left[\sum_{t=1}^T d(a_{t-1}, a_t) + c_t(a_t)\right] &= \E\left[ \sum_{t=1}^T d(a_{t-1}, a_t) + c_t(a_t) + u_{x_t}(a_t) - u_{x_t}(a_t)\right] \\
    &= \E\left[ \sum_{t=1}^T B^{c_t}u_{x_t}(a_{t-1}) - u_{x_t}(a_t)\right] \\
    &= \E\left[ \sum_{t=1}^T B^{c_t}u_{x_t}(a_{t-1}) - u_{x_{t-1}}(a_{t-1})\right] + \E[u_{x_0}(a_0) - u_{x_T}(a_T)] \\ 
    &= \E\left[ \sum_{t=1}^T (\bar{B}u)_{x_{t-1}}(a_{t-1}) - u_{x_{t-1}}(a_{t-1})\right] + \E[u_{x_0}(a_0) - u_{x_T}(a_T)] \\
    &\leq \E\left[ \sum_{t=1}^T \max(v_{x_{t-1}}) \right] + \E[u_{x_0}(a_0) - u_{x_T}(a_T)]
    \end{align*}

Here, in the second equality we used the definitions of $B^{c_t}u$ and $a_t$, in the penultimate step we used the definition of $\bar{B}$, and in the last step we used the definition of $v$. Dividing by $T$ and letting $T\rightarrow \infty$, the second term vanishes, and we  obtain 
\begin{equation}
\label{eq:bounding-g-piu-Markov}
g(\pi_u) \leq \sum_{x \in \mathcal{X}} p_x \max(v_x).
\end{equation}

 Let $o$ be the trajectory induced by an optimal policy $\pi^*$, recursively defined by $o_t \coloneqq \pi^*(o_{t-1}, c_t, x_t)$. Similarly to the above, we show that the expected cost of policy $\pi^*$ for the first $T$ time steps is 
 
 \begin{align*}
    \E \left[\sum_{t=1}^T d(o_{t-1}, o_t) + c_t(o_t)\right] &= \E\left[ \sum_{t=1}^T d(o_{t-1}, o_t) + c_t(o_t) + u_{x_t}(o_t) - u_{x_t}(o_t)\right] \\
    &\geq \E\left[ \sum_{t=1}^T B^{c_t}u_{x_t}(o_{t-1}) - u_{x_t}(o_t)\right] \\
     &= \E\left[ \sum_{t=1}^T B^{c_t}u_{x_t}(o_{t-1}) - u_{x_{t-1}}(o_{t-1}) \right] + \E[u_{x_0}(o_0) - u_{x_T}(o_T)] \\
      &= \E\left[ \sum_{t=1}^T (\bar{B}u)_{x_{t-1}}(o_{t-1}) - u_{x_{t-1}}(o_{t-1})\right] + \E[u_{x_0}(o_0) - u_{x_T}(o_T)] \\
      &\geq  \E\left[ \sum_{t=1}^T \min(v_{x_{t-1}}) \right] + \E[u_{x_0}(o_0) - u_{x_T}(o_T)].
    \end{align*}

Dividing by $T$ and letting $T\rightarrow \infty$, we  obtain 
\begin{equation}
\label{eq:bounding-g*-Markov}
 g(\pi^*) \geq\sum_{x \in \mathcal{X}} p_x \min(v_x).
\end{equation}

Putting \eqref{eq:bounding-g-piu-Markov} and \eqref{eq:bounding-g*-Markov} together, we obtain 
\[
\sum_{x \in \mathcal{X}} p_x \min(v_x) \leq g(\pi^*) \leq g(\pi_u) \leq \sum_{x \in \mathcal{X}} p_x \max(v_x).
\]
By rearranging, applying triangle inequality, and using the definition of $\|\cdot\|$,
\begin{align*}
g(\pi_u) - g^* \leq \sum_{x \in \mathcal{X}} p_x \|v_x\|_\SP &= \sum_{x \in \mathcal{X}} p_x \|(\overline{B}u)_x - u_x\|_\SP \\
&\leq \sum_{x \in \mathcal{X}} p_x \|(\widetilde{B}u)_x - u_x\|_\SP + \sum_{x \in \mathcal{X}} p_x \|(\overline{B}u)_x - (\widetilde{B}u)_x\|_\SP \\
&\leq \|\widetilde{B}u - u\| + \sum_{x \in \mathcal{X}} p_x \|(\overline{B}u)_x - (\widetilde{B}u)_x\|_\SP.\qedhere
\end{align*}
\end{proof}
\subsection{Uniform estimation of the Bellman operator}
\label{app:concentration}

We now control the error caused by replacing the expected Bellman
operator with its empirical counterpart. For the remainder of this
appendix, write
\[
    e_x(u)\coloneqq\|(\overline Bu)_x-(\widetilde Bu)_x\|_{\SP}.
\]
By Lemma~\ref{lemma:to-conclude-Markov}, it suffices to show that,
with probability at least $1-\delta$,
\[
    \sum_{x\in\mathcal X} p_x e_x(u)\le \frac{\varepsilon}{2}
    \qquad\text{for every }u\in C.
\]
Uniformity over $C$ ensures that this bound applies to the
data-dependent Halpern iterates, while weighting by $p_x$ allows
less accurate estimates at rarely visited states.
We first prove the bound for the two oracle models, developing
estimates that we will subsequently reuse for learning from
a single trajectory.

\subsubsection{Learning from oracle access}
\label{app:oracle-concentration}
Recall the two types of oracles we consider: With \emph{conditional oracles}, the learner chooses a state $x$ and
receives an independent pair $(x',c)\sim\mathcal D_x$.
With a \emph{stationary oracle}, each call independently returns a
triple $(x,x',c)$ by first drawing $x\sim p$, where $p$ is the stationary distribution of the Markov chain, and then
$(x',c)\sim\mathcal D_x$. 

In both cases, we denote by $N$ the total number of oracle calls, and by $(x_i, x'_i, c_i)_{i=1}^N$ the list of triples obtained. Let 
\[
    I_x\coloneqq\{(x'_i, c_i):x_i=x\}, \qquad N_x\coloneqq |I_x|.
\]

Also, let $m\coloneqq\vert \mathcal{X}\vert$, and set
\[
    \Gamma
    \coloneqq nm\log\left(\left\lceil\frac{64D}{\varepsilon}\right\rceil+1\right)
      +\log\frac{4nm}{\delta}.
\]

\begin{claim}[Uniform estimation for oracle samples]
\label{cl:uniform-estimation-oracles}
Suppose that the training data are obtained either by taking a fixed
number of independent samples from each conditional oracle, or by making
a fixed number of independent stationary-oracle calls. With probability
at least $1-\delta/2$, simultaneously for every $u\in C$ and every $x$
with $N_x>0$,
\begin{equation}
    e_x(u)\le D\sqrt{\frac{2\Gamma}{N_x}}+\frac\varepsilon8.
    \label{eq:app-oracle-estimation}
\end{equation}
\end{claim}
\begin{proof}

Let
$
H \coloneqq ([0,D]^M)^\mathcal{X},
$
and note that 
it suffices to prove the claim uniformly over $u \in H$. For $x' \in \mathcal{X}$, $u \in H$, and $c\colon M \rightarrow \reals_{\geq 0} \cup \{\infty\}$ with $\min(c)<\infty$, define
\[
\Phi_{x',c}(u) \coloneqq B^c u_{x'} - \min(B^c u_{x'}).
\]
Each $B^cu_{x'}$ is $1$-Lipschitz on $M$ and hence
has span at most $D$, so $0\le\Phi_{x',c}(u)(s)\le D$.

Let $\eta\coloneqq\varepsilon/64$, and let $\mathcal N$ be an $\eta$-net of $H$ with respect to $\|\cdot\|_{\infty}$. This means that for each $u\in H$, there exists $w\in\mathcal{N}$ such that $\vert u-w\vert_\infty \leq \eta$. We can choose $\mathcal{N}$ such that
\[
    |\mathcal N|
       \le\left(\left\lceil\frac D\eta\right\rceil+1\right)^{nm}
       =\left(\left\lceil\frac{64D}{\varepsilon}\right\rceil+1\right)^{nm}.
\]

For the conditional oracles, the $x_1, \dots, x_N$ are fixed in
advance by the learner. For the stationary oracle, condition on $x_1,\ldots,x_N$. In either case, the counts $N_x$ are
fixed and, for each $x$, the pairs
in $I_x$ are independent samples from $\mathcal D_x$. For fixed $w\in\mathcal N$, $s\in M$, and $x\in\mathcal{X}$ with $N_x>0$,
Hoeffding's inequality gives, conditionally on $x_1,\dots,x_N$,
\[
    \Prob\left(
       \left|
    \frac{1}{N_x} \sum_{(x',c) \in I_x} \Phi_{x', c}(w)(s)
    -
    \mathbb E_{(x',c) \sim \mathcal D_x}[\Phi_{x',c}(w)(s)]
\right|\
       >D\sqrt{\frac{\Gamma}{2N_x}}
    \right)\le2e^{-\Gamma}.
\]
A union bound over such $w,x,s$ gives a conditional failure probability
at most
\[
    2|\mathcal N|mn e^{-\Gamma}\le\frac\delta2.
\]
This holds for every realization of $x_1,\dots,x_N$, so the same bound holds unconditionally.

On this event, fix $u\in H$ and choose $w\in\mathcal N$ with
$\|u-w\|_\infty\le\eta$.
We have
\begin{equation}
    \|\Phi_{x',c}(u)-\Phi_{x',c}(w)\|_\infty
       \le2\|u-w\|_\infty\leq 2\eta,
    \label{eq:app-normalized-output-lipschitz}
\end{equation}
and therefore
\[
       \left\|
    \frac{1}{N_x} \sum_{(x',c) \in I_x} \Phi_{x', c}(u)
    -
    \mathbb E_{(x',c) \sim \mathcal D_x}[\Phi_{x',c}(u)]
\right\|_\infty
       \leq D\sqrt{\frac{\Gamma}{2N_x}} + 4\eta.
\]

Since
$\|v\|_{\SP}\le2\|v\|_\infty$, it follows that
\begin{equation}
    \left\|\frac{1}{N_x} \sum_{(x',c) \in I_x} \Phi_{x', c}(u)
    -
    \mathbb E_{(x',c) \sim \mathcal D_x}[\Phi_{x',c}(u)]\right\|_{\SP}
       \le D\sqrt{\frac{2\Gamma}{N_x}}+8\eta
       =D\sqrt{\frac{2\Gamma}{N_x}}+\frac\varepsilon8.
    %\label{eq:app-oracle-estimation}
\end{equation}
Note that $\frac{1}{N_x} \sum_{(x',c) \in I_x} \Phi_{x', c}(u)$ and $\mathbb E_{(x',c) \sim \mathcal D_x}[\Phi_{x',c}(u)]$ are equivalent to $(\widetilde{B}u)_x$ and $(\overline{B}u)_x$, respectively, up to an additive constant, and hence taking the span seminorm gives exactly 
$e_x(u)$.
\end{proof}

The following lemma provides sufficient conditions for bounding $\sum_x p_x e_x(u)$. It does not make any assumption on how the observations $(I_x)_{x\in\mathcal{X}}$ are obtained.

\begin{lemma}[Sufficient visit counts]
\label{lem:trajectory-sufficient-counts}
Let $A>0$.
Suppose that, simultaneously for every $u\in C$ and every $x$ with
$N_x>0$,
\[
    e_x(u)\le D\sqrt{\frac{2A}{N_x}}+\frac\varepsilon8.
\]
Set $\rho\coloneqq\varepsilon/(16Dm)$, and suppose that
\begin{equation}
    N_x\ge Np_x/4
       \qquad\text{whenever }p_x>\rho.
    \label{eq:app-sufficient-counts}
\end{equation}
If
\begin{equation}
    N\ge\frac{128D^2mA}{\varepsilon^2},
    \label{eq:app-estimation-budget}
\end{equation}
then $\sum_xp_xe_x(u)\le\varepsilon/2$ for all $u\in C$.
\end{lemma}

\begin{proof}
Fix $u\in C$. Note that $e_x(u) \leq 2D$ for all $x\in\mathcal{X}$, because both $(\overline{B}u)_x$ and $(\widetilde{B}u)_x$ have span at most $D$. Therefore, the states with
$p_x\le\rho$ contribute at most
\[
    \sum_{x:p_x\le\rho}p_xe_x(u)
       \le2Dm\rho=\frac\varepsilon8.
\]
For the remaining states, the estimation bound and
\eqref{eq:app-sufficient-counts} imply
\begin{align*}
    \sum_{x:p_x>\rho}p_xe_x(u)
    &\le D\sqrt{2A}\sum_{x:p_x>\rho}\frac{p_x}{\sqrt{N_x}}
          +\frac\varepsilon8\sum_{x:p_x>\rho}p_x\\
    &\le D\sqrt{\frac{8A}{N}}\sum_{x:p_x>\rho}\sqrt{p_x}
          +\frac\varepsilon8\\
    &\le D\sqrt{\frac{8mA}{N}}+\frac\varepsilon8,
\end{align*}
where the last step uses Cauchy--Schwarz and $\sum_xp_x=1$.
Together with~\eqref{eq:app-estimation-budget}, this gives
\begin{equation*}
    \sum_xp_xe_x(u)
       \le\frac\varepsilon8+\frac\varepsilon8+D\sqrt{\frac{8mA}{N}}
       \le\frac\varepsilon2.
    \label{eq:app-weighted-estimation-bound}\qedhere
\end{equation*}
\end{proof}

We are now ready to prove Theorem \ref{th:culminating-Markov}, restated below.

\theoremMarkov*
\begin{proof}
We treat the two types of oracles separately. 

\textbf{Conditional oracles.}
We make $N_x \coloneqq O\left(\frac{D^2 \card{\mathcal{X}}}{{\varepsilon}^2} \left(n \log \frac{D}{\varepsilon}
        + \log \frac{n \vert \mathcal{X} \vert}{\delta}\right) \right)$ calls to the conditional oracle for each state $x\in\mathcal{X}$. By Claim \ref{cl:uniform-estimation-oracles} and our choice of $N_x$, with probability at least $1-\delta/2$,
\[
    e_x(u)\le D\sqrt{\frac{2\Gamma}{N_x}}+\frac\varepsilon8
       \le\frac\varepsilon4
    \qquad\text{for every }u\in C\text{ and }x\in\mathcal X.
\]
Combining this with Lemma \ref{lemma:to-conclude-Markov} concludes the proof.

\textbf{Stationary oracle.}
Here $N_x$ is binomial with parameters $N$ and $p_x$, so the Chernoff bound yields
\[
    \Prob(N_x<Np_x/2)\le e^{-Np_x/8}.
\]
Set $\rho\coloneqq\varepsilon/(16Dm)$. A union bound shows that, with
probability at least $1-\delta/2$, $N_x\ge Np_x/2$ for every state
with $p_x>\rho$, provided
\begin{equation}
    N\ge\frac8\rho\log\frac{2m}{\delta}
       =\frac{128Dm}{\varepsilon}\log\frac{2m}{\delta}.
    \label{eq:app-iid-count-budget}
\end{equation}
Our choice of $N$ satisfies this because $D/\varepsilon\ge1$ and
$\Gamma\ge\log(2m/\delta)$. On this count event,
\eqref{eq:app-sufficient-counts} holds. Moreover, the definition of $N$
ensures~\eqref{eq:app-estimation-budget} with $A=\Gamma$.
Combining Lemma~\ref{lem:trajectory-sufficient-counts} and Lemma ~\ref{lemma:to-conclude-Markov} concludes the proof.
\end{proof}

\subsubsection{Learning from an observed trajectory}
\label{app:trajectory-concentration}
\label{app:trajectory-theorem}

We consider the trajectory $x_0,(x_1,c_1),\ldots,(x_N,c_N)$ from an
arbitrary initial state. Thus $N_x$ counts visits to
$x$ at times $0,\ldots,N-1$.

Define
\begin{equation}
    \Gamma_N\coloneqq\Gamma+\log N
    =nm\log\left(\left\lceil\frac{64D}{\varepsilon}\right\rceil+1\right)
      +\log\frac{4nmN}{\delta}.
    \label{eq:trajectory-concentration-factor}
\end{equation}

\begin{lemma}[Uniform estimation along a trajectory]
\label{lem:trajectory-uniform-estimation}
For $N$ transitions from any initial Markov state, with probability
at least $1-\delta/2$, simultaneously for every $u\in C$
and every $x$ with $N_x>0$,
\[
    e_x(u)\le D\sqrt{\frac{2\Gamma_N}{N_x}}+\frac\varepsilon8.
\]
\end{lemma}

\begin{proof}
We mostly reuse the analysis from Claim \ref{cl:uniform-estimation-oracles}, with one caveat: conditioned on $x_1, \dots, x_N$, 
the observations $I_x$ are no longer independent samples from $\mathcal{D}_x$.

To handle this, we proceed as follows. For each $x\in\mathcal X$, generate an infinite independent sequence
$Y_{x,1},Y_{x,2},\ldots\sim\mathcal D_x$, independently across states.
When the current state is $x$, reveal the next unused
element of this sequence. Its conditional distribution given the
past is $\mathcal D_x$, so this generates the required Markov request
process. The observed samples from $x$ are exactly
$Y_{x,1},\ldots,Y_{x,N_x}$.

Apply the Hoeffding bound from the proof of
Claim~\ref{cl:uniform-estimation-oracles} to every prefix length
$1\le\ell\le N$ of the sequence $Y_{x,1},Y_{x,2},\ldots$, with $\Gamma_N$ in place of $\Gamma$.
For fixed $w\in\mathcal N$, $x$, $s$, and $\ell$, the deviation
threshold is $D\sqrt{\Gamma_N/(2\ell)}$ and the failure probability
is at most $2e^{-\Gamma_N}$. The additional union bound over $\ell$
costs a factor $N$, which is exactly absorbed by
$\Gamma_N=\Gamma+\log N$:
\[
    2|\mathcal N|mnN e^{-\Gamma_N}\le\frac\delta2.
\]
The net-extension and translation arguments giving
\eqref{eq:app-oracle-estimation} apply unchanged. On a single event
of probability at least $1-\delta/2$, for every $u\in C$,
$x\in\mathcal X$, and $1\le\ell\le N$,
\begin{equation}
    \left\|\frac1\ell\sum_{j=1}^{\ell}\Phi_{Y_{x,j}}(u)
          -\mathbb E_{Y\sim\mathcal D_x}[\Phi_Y(u)]\right\|_{\SP}
       \le D\sqrt{\frac{2\Gamma_N}{\ell}}+\frac\varepsilon8.
    \label{eq:app-all-prefixes}
\end{equation}
For each $x\in\mathcal{X}$ with $N_x>0$, take $\ell=N_x$. The estimate holds
simultaneously for every $\ell$, so substituting the random $N_x$
is valid without conditioning on it. The left-hand side equals
$e_x(u)$, as in \eqref{eq:app-oracle-estimation}.
\end{proof}

Let \[d(t)\coloneqq\max_x\|P^t(x,\cdot)-p\|_{\mathrm{TV}},\]
where $P$ denotes the transition matrix of the Markov chain and $\|\cdot\|_{\mathrm{TV}}$ denotes the total variation distance, defined by 
\[
    \|\mu-\nu\|_{\mathrm{TV}}
       \coloneqq\frac12\sum_{x\in\mathcal X}|\mu_x-\nu_x|.
\]
Then, the mixing time of the chain is defined by
\begin{equation}
    \tau\coloneqq\min\left\{t\ge1:
       d(t)\le\frac18\right\}.
    \label{eq:trajectory-mixing-time}
\end{equation}

We next establish the necessary visit-count bound. 
\begin{lemma}[Visit counts]
\label{lem:trajectory-visit-counts}
There is an absolute constant $K_0>0$ such that, for every
$\rho\in(0,1]$, $\delta\in(0,1)$, and integer $N$ satisfying
\begin{equation}
    \label{eq:app-trajectory-count-budget}
    N\ge K_0\frac{\tau}{\rho}\log\frac{8m}{\delta},
\end{equation}

with probability at least $1-\delta/2$,
\[
    N_x=|\{0\le t<N:x_t=x\}|\ge\frac{Np_x}{4}
    \qquad\text{for every }x\text{ with }p_x\ge\rho.
\]
\end{lemma}

\begin{proof}
The standard mixing bound~\cite{LP17} gives $d(t)\le2^{-\lfloor t/\tau\rfloor}$. Set $b\coloneqq\tau\lceil\log_2(4/\delta)\rceil$, so that $d(b)\leq \delta/4$, and let $R \coloneqq N-b$. By maximal coupling~\cite[Proposition~4.7]{LP17},
$(x_b,\ldots,x_{N-1})$ can be coupled to a stationary trajectory
$(\widehat x_0,\ldots,\widehat x_{R-1})$ so that 
\[
\Prob\left((x_b,\ldots,x_{N-1}) = (\widehat x_0,\ldots,\widehat x_{R-1})\right) \geq 1-d(b)\ge1-\delta/4.
\]

On coupling success, 
\[
N_x = \sum_{t=0}^{N-1} \mathbf 1_{\{x_t=x\}} \geq \sum_{t=b}^{N-1} \mathbf1_{\{x_t=x\}} = \sum_{t=0}^{R-1} \mathbf1_{\{\widehat x_t=x\}}.
\]

For a fixed state $x$, apply the lower-tail bound of
Chung--Lam--Liu--Mitzenmacher~\cite[Theorem~3]{CLLM12}. Since the auxiliary chain starts
in stationarity, this gives
\[
    \Prob\!\left(\sum_{t=0}^{R-1}\mathbf 1_{\{\widehat x_t=x\}}
                   \le\frac{Rp_x}{2}\right)
    \le c_0\exp\!\left(-\frac{Rp_x}{300\tau}\right),
\]
where $c_0$ is an absolute constant.

We will choose $N$ to ensure that $R\geq N/2$, and hence $Rp_x/2\ge Np_x/4$. By union bound,
\[
    \Prob\!\left(\exists x:p_x\ge\rho,\ N_x<\frac{Np_x}{4}\right)
    \le\frac{\delta}{4}
       +c_0m\exp\!\left(-\frac{N\rho}{600\tau}\right)
    \le\frac{\delta}{2}.
\]
Setting $N \geq \max\{\lceil600\frac{\tau}{\rho}\log\frac{4mc_0}{\delta}\rceil, 2b\}$ completes the proof.
\end{proof}

We are now ready to prove Theorem~\ref{th:culminating-Markov-trajectory}, restated below.
\theoremTrajectory*
\begin{proof}
Set
\begin{equation}
    N\coloneqq\left\lceil
        C_0\left(
            \frac{D^2m}{\varepsilon^2}\Gamma
            +\frac{\tau Dm}{\varepsilon}\log\frac{8m}{\delta}
        \right)
    \right\rceil,
    \label{eq:app-explicit-trajectory-budget}
\end{equation}
for a sufficiently large absolute constant $C_0$.

We verify the prerequisites needed to apply
Lemma~\ref{lem:trajectory-sufficient-counts}. The first term in the definition ensures that, for large enough $C_0$,
\[
    N\ge128\frac{D^2m}{\varepsilon^2}(\Gamma + \log N),
\]
which is precisely~\eqref{eq:app-estimation-budget} with $A=\Gamma_N$.

By setting $\rho^{-1}\coloneqq16Dm/\varepsilon$,  the second term
in~\eqref{eq:app-explicit-trajectory-budget} ensures
\[
    N\ge K_0\frac{\tau}{\rho}\log\frac{8m}{\delta},
\]
which satisfies~\eqref{eq:app-trajectory-count-budget},
for sufficiently large $C_0$.

Lemma~\ref{lem:trajectory-visit-counts} gives, with probability at least $1-\delta/2$, $N_x\ge{Np_x}/{4}$ for all $x$ with $p_x\ge\rho$.
Lemma~\ref{lem:trajectory-uniform-estimation} gives the required
uniform estimation bound, with $A=\Gamma_N$, with probability at
least $1-\delta/2$. Taking a union bound and applying 
Lemma~\ref{lem:trajectory-sufficient-counts} combined with Lemma ~\ref{lemma:to-conclude-Markov} yields the conclusion.
\end{proof}

\section{Proof of Halpern residual bound (Theorem \ref{thm:halpern-rate2})} \label{sec:halpern-bound-proof}

This section contains a proof of Theorem \ref{thm:halpern-rate2}, restated here for convenience. The proof is very similar to the one given in \cite{SaSh17}.

\Halpernbound*

\begin{proof}

    To bound $\|z_k-Fz_k\|$, we decompose it into 
    \begin{equation}
    \label{eq:decompose}
    \|z_k-Fz_k\| \leq  \|z_k-z_{k+1}\| + \|z_{k+1}-Fz_k\|.
    \end{equation}

    The second term is easy to bound:
    \begin{equation}
    \label{eq:bound-second-term}
    \begin{aligned}
    \|z_{k+1}-Fz_k\| &= \|  (1-\alpha_{k+1})F z_k - Fz_k \| \\
    &= \alpha_{k+1}\| F z_k \| \\ 
    &\leq \alpha_{k+1} \diam(C)\\
    &= \frac{2}{k+3} \diam(C).
    \end{aligned}
    \end{equation}

    To bound the first term, we proceed as follows:
    \begin{align*}
    \|z_k-z_{k+1}\| &= \| (1-\alpha_k)F z_{k-1} - (1-\alpha_{k+1})F z_k \| \\
    &\leq (1-\alpha_k) \| F z_{k-1} -  F z_k\| + (\alpha_k - \alpha_{k+1}) \| Fz_k \| .
    \end{align*}

    Since $Fz_k\in C$, we have $\| Fz_k \| \leq \diam(C)$. Since $F$ is non-expansive, 
    \[ \| F z_{k-1} -  F z_k\| \leq \| z_{k-1} - z_k \|.\]

    Putting these together, and denoting $\|z_k-z_{k+1}\|$ by $l_k$, we get
    \[
    l_k \leq (1 - \alpha_k) l_{k-1} + (\alpha_k - \alpha_{k+1})\diam(C).
    \]
    Unfolding this, and noting that $l_0 \leq \diam(C)$, we get
    \begin{align*}
        l_k \leq \sum_{i=2}^{k+1} \left((1-\alpha_k) \dots (1 - \alpha_i)\right) (\alpha_{i-1} - \alpha_i) \diam(C) +  (1-\alpha_k) \dots (1-\alpha_1) \diam(C).
    \end{align*}

    Substituting $\alpha_i = \frac{2}{i+2}$, we get that 
    \[
    (1-\alpha_k) \dots (1 - \alpha_i) = \frac{k}{k+2} \dots \frac{i}{i+2} =  \frac{i(i+1)}{(k+1)(k+2)}
    \]
    and 
    \[
    \alpha_{i-1} - \alpha_i = \frac{2}{(i+1)(i+2)}.
    \]
    Therefore, 
    \begin{align*}
    l_k &\leq \left(\sum_{i=2}^{k+1} \frac{i(i+1)}{(k+1)(k+2)} \cdot \frac{2}{(i+1)(i+2)} + \frac{2}{(k+1)(k+2)}\right) \diam(C)
    \end{align*}
    We can bound each term by $\frac{2}{(k+1)(k+2)}$ to obtain
    \begin{align*}
    l_k \leq (k+2) \frac{2}{(k+1)(k+2)}\diam(C) = \frac{2}{k+1}\diam(C).
    \end{align*}

    Using this together with \eqref{eq:decompose} and \eqref{eq:bound-second-term}, we get the desired result.
\end{proof}

%\newpage
%\input{checklist.tex}

\end{document}